\documentclass[aps,twocolumn,showpacs]{revtex4}
\usepackage{amsmath}
\usepackage{epsfig}
\usepackage{amssymb}
\begin{document}

\title{Knot and Gauge Theory}

\newcommand*{\NJNU}{Department of Physics, Nanjing Normal University, Nanjing, Jiangsu 210097, China}\affiliation{\NJNU}

\author{Jing Zhou}\email{171001005@stu.njnu.edu.cn}\affiliation{\NJNU}
\author{Jialun Ping}\email{jlping@njnu.edu.cn}\affiliation{\NJNU}

\begin{abstract}
It has been argued based on electric-magnetic duality
that the Jones polynomial of a knot in three dimensions can be computed by counting the
solutions of certain gauge theory equations in four-dimension. And the Euler characteristic of Khovanov homology 
is the Jones polynomial which corresponds to the partition function of twisted $N=4$ super Yang-Mills theory. 
Moreover, Lee-Yang type phase transition is found in the topological twisted super Yang-Mills theory.
\end{abstract}


\maketitle

\section{\label{sec:introduction}Introduction}
The Jones polynomial\cite{Jones:1985dw,Jones:1987dy} is a celebrated knot invariant in the three-dimensional space, 
it is discovered by Jones base on the work of von Neumann algebras. Actually, many generalizations of the 
Jones polynomial were discovered after Jones's work. And it has multiple relations to many aspects of mathematical 
physics, which including representations of braid groups, two dimensional conformal field theory, statistical mechanics 
and three-dimensional Chern-Simons gauge theory\cite{Freyd:1985dx,Turaev:1988eb,Jones:1989ed,Akutsu:1987bc,Witten:1988hf}. 


Khovanov homology \cite{WItten:2011pz} is considered to be a topological theory in four-dimension super Yang-Mills 
field theory. In this theory, a knot is viewed as an object in three dimensional space and the invariant associated 
to a knot is a vector space instead of a number. The relation between the Khovanov homology and the knot is that 
the four-dimensional theory associated to Khovanov homology, when compactified on a circle, reduces to the 
three-dimensional theory that yields the Jones polynomial.

Lee-Yang theory is also discussed in the work. Actually, Lee-Yang~\cite{Lee:1952ig,Yang:1952be} phase transition 
has gone beyond the statistical physics. Maloney and Witten~\cite{Maloney:2007ud} shown that the Hawking-Page 
transition~\cite{Hawking:1982dh} in $AdS_{3}$ space is Lee-Yang type, while the original Lee-Yang phase transition 
is only for two dimensional Ising model. This is an important reason that we argue the Lee-Yang type phase 
transition happens in the super Yang-Mills field theory. Next, let us interpret the Lee-Yang phase transition 
in the $AdS_{3}$ space in more detail.

The Hawking-Page transition can be seen from Lee-Yang condensation of zeros in the partition function for 
$k\rightarrow\infty$. Actually, the partition function $Z(\tau)$ of three dimensional gravity is a modular 
function which computes at fixed temperature Im$\tau$ and angular potential Re$\tau$. In the limit of 
infinite volume, these zeroes condense along the phase boundaries. So, it gives rise to phase transition. 
The analog of the infinite volume limit for the partition function $Z(\tau)$ is $k\rightarrow\infty$ because 
as we know $k=\ell / 16 G$ implies that k directly proportional to the $AdS_{3}$ radius. In this case, 
the partition function $Z(\tau)$ is not analytic which corresponding to the occurrence of phase transition.

Our aim here, however, is to understand not phase transition in $AdS_{3}$ space but phase transition in 
super Yang-Mills field theory. In the following work, we show that it may shed light on this issue.

\section{The Jones Polynomial and Topological Twisted Super Yang-Mills Theory }
The Chern-Simons action for a gauge theory with gauge group $G$ and gauge field $A$ on an oriented 
three-manifold $W$ can be write as~\cite{Witten:2010cx,Gopakumar:1998ki},
\begin{equation}
 CS(A)= \frac{k}{4\pi}\int_{W}Tr\left(dA\wedge A+\frac{2}{3}A\wedge A\wedge A\right).
 \end{equation}
Here $k$ is an integer for topological reasons. Then one can find that $CS(A)$ is gauge-invariant. 
Now, the Feynman path integral is an integral over the infinite-dimensional space of connections $A$. 
In fact, this is a basic construction in the quantum field theory. So, we can write as,
\begin{equation}
\int DA\exp\left(iCS\right).
\end{equation}
If making use of the loop of the connection $A$ around $K$, then we obtain a knot which it is an embedded 
oriented loop $K \subset M$. If picking an irreducible representation $R$ of $K$, then one can associate 
an observable which is the trace of Wilson loop operator,
\begin{equation}
 W\left(K,R\right)= Tr P\exp\int_{K} A .
\end{equation}
Once one does this, one can we define a natural invariant of the pair ($W$, $K$), So the partition function 
can write as~\cite{Witten:2011zz},
\begin{equation}
 Z\left(M; K\right)= \int DA\exp\left(iCS\right)\prod_{i=1}^{r} W\left(K_{i},R\right).
\end{equation}
This is a topological invariant of the knot $K$ in the three manifold $W$, which depends only on $G$, $R$ and $K$. 
Then, Witten made use of the topological invariance of the theory to solve Chern-Simons topological theory on 
three manifold $W$ with collection of knots, which is a suitably normalized Chern-Simons partition function,
\begin{equation}
  \langle\mathcal{W}(K, R)\rangle= 
  \frac{\int DA\exp\left(iCS\right)\prod_{i=1}^{r} W\left(K_{i}, R\right)}{\int DA\exp\left(iCS\right)}
\end{equation}
equals to the Jones polynomial,
\begin{equation}
\left\langle\mathcal{W}(K, R)\right\rangle= J_{K}(q).
\end{equation}
In fact, this knot invariant of a simple Lie group $G$ on a three manifold $W$ can be computed by counting 
solutions of a certain system of elliptic partial differential equations, with gauge group the dual group $G$, 
on the four-manifold $M_{4}=W \times \mathbb{R}_{+}$, where $W$ is a three manifold and $\mathbb{R}_{+}$ is 
the half line $y \geq 0$. The equations are the KW equations~\cite{Gaiotto:2011nm,Kapustin:2006pk}.
\begin{equation}
\begin{aligned}\left(F-\phi \wedge \phi+\mathrm{t} \mathrm{d}_{A} \phi\right)^{+} &=0 \\
\left(F-\phi \wedge \phi-\mathrm{t}^{-1} \mathrm{d}_{A} \phi\right)^{-} &=0 \\ 
\mathrm{d}_{A} \star \phi &=0 \end{aligned}
\end{equation}
Where $F=\mathrm{d} A+A \wedge A$ is the Yang-Mills field strength, $\star$ is the Hodge star operator, 
and t is a real parameter. In fact, the the KW equations arise in twisted $N = 4$ super Yang-Mills theory in 
four-dimension, known as geometric Langlands duality~\cite{Kapustin:2006pk}. And it can be used to describe 
the Jones polynomial and similar invariants of knot. Or, in other words, the quantum invariants of a knot in 
$\mathbb{R}^{3}$ can be computed from a path integral of twisted $N =4$ super Yang-Mills theory in four-dimension, 
with the boundary condition $\mathbb{R}^{3} \times\{0\}$.

In fact, the Jones polynomial can expand as an Euler characteristic as a trace in which bosonic and fermionic 
states cancel, in the invariantly defined cohomology $H$, which can write as~\cite{WItten:2011pz},
\begin{equation}
J_{k}(q)=\chi=\operatorname{Tr}_{H}(-1)^{F} q^{P}.
\end{equation}
On the other hand, the partition function of twisted $N=4$ super Yang-Mills is Euler characteristic~\cite{Vafa:1994tf},
\begin{equation}
Z_{twist}=\chi(\mathcal{W})
\end{equation}
with $W$ is the space of solution of KW equation. So, the Euler characteristic of $M_{4}=\mathbb{R}^{3} \times\{0\}$ 
is only contributed by $\mathbb{R}^{3}$. Then we can write as,
$$J_{k}(q)= Z_{twist}$$
or, in other words, the partition function on the boundary can expand as Jones polynomial.  And this knot invariant 
used to consider as a Laurent polynomial in $q=\exp \left(\frac{2 \pi i}{k+2}\right)$.

\section{Lee-Yang Type Phase Transition of Super Yang-Mills Theory }
Topological field theory has been important for establishing symmetries and dualities in field theory and 
string theory~\cite{Witten:1988ze,Bershadsky:1993cx,Manschot:2017xcr}. A generalization of electric-magnetic 
duality to Yang-Mills theory, known as $S$-duality, acts naturally on Vafa-Witten theory~\cite{Vafa:1994tf}. 
This duality, proposed by Montonen and Olive~\cite{Montonen:1977sn}, states that Yang-Mills theory with gauge group 
$G$ and complexified coupling constant
\begin{equation}
\tau=\frac{\theta}{2 \pi}+\frac{4 \pi i}{g^{2}}
\end{equation}
has a dual description as Yang-Mills theory, whose gauge group is the Langlands dual group and with inverse 
coupling constant $-1 / \tau$ . Together with the periodicity of the $\theta$-angle, this generates the 
$S L(2, \mathbb{Z})$ $S$-duality group,
\begin{equation}
Z\left(\frac{a \tau+b}{c \tau+d}\right) \sim Z(\tau), \quad \left( \begin{array}{ll}{a} & {b} \\ 
{c} & {d}\end{array}\right) \in S L(2, \mathbb{Z})
\end{equation}
The Vafa-Witten twist of $N = 4$ super symmetric Yang-Mills theory with gauge group $SU(N)$ contains a commuting 
BRST-like operator $Q$. For a suitable $I$, the topologically twisted action of Vafa-Witten theory can be expressed 
as a $Q$-exact term ${Q, I}$, plus a term multiplying the complexified coupling constant $\tau$:
\begin{equation}
\mathcal{S}_{\text { twisted }}=\{\mathcal{Q}, I\}-2 \pi i \tau(n-\Delta),
\end{equation}
where $n$ denotes the instanton number,
\begin{equation}
n=\frac{1}{8 \pi^{2}} \int_{M} \operatorname{Tr} F \wedge F
\end{equation}
In fact, the partition function of the topological twisted super Yang-Mills field theory depend only on 
$\tau$ and the gauge group $SU(N)$ chosen, which is a modular form~\cite{Vafa:1994tf,Labastida:1999ij,Witten:1998wy},
\begin{equation}
\begin{array}{c}{Z_{S U(N)}(-1 / \tau)=\pm N^{-1+b_{1}-\frac{b_{2}}{2}}\left(\frac{\tau}{i}\right)^{\frac{w}{2}} 
 Z_{S U(N) / \mathbf{Z}_{N}}(\tau)} \\ {=\pm N^{-\chi / 2}\left(\frac{\tau}{i}\right)^{\frac{w}{2}} Z_{S U(N) 
 / \mathbf{Z}_{N}}(\tau)}
\end{array}
\end{equation}
Where $\chi$ is Euler characteristics, $w$ is the modular weight. For any finite value of $N$, the partition function 
$Z(\tau)$ is smooth as a function of $\tau$, but for large $N$ limit, as $N \rightarrow \infty$, the function becomes
non-smooth~\cite{Maloney:2007ud}. Or, in other words, the function is not a analytic function now. The original idea 
of Lee and Yang is that although a system in finite volume can have no phase transition, its partition function, 
depending on the complexified thermodynamic variables, can have zeroes. Then, in the infinite volume limit, the zeroes 
become more numerous and may become dense. Then, a true phase transition can emerge. In our problem, the limit 
$N \rightarrow \infty$, is analogous to a thermodynamic limit.

Now, we can find that the partition function $Z(\tau)$ is topological invariant(Euler characteristic). 
In mathematics, we can write,
\begin{equation}
\chi=\sum_{i=1}^{d}(-1)^{i} \operatorname{dim} H^{i}.
\end{equation}
If only considering field coefficients, then the cohomology group $H^{i}$ is a vector space. One should note that 
the partition function $Z(\tau)$ equal to the Euler characteristic. Then, in the large $N$ limit, phase transition 
happens. Or in other words, the partition function $Z(\tau)$ (Euler characteristic) is not an analytic function. 
So the cohomology group is not analytic too. And this argument may important in mathematics.

\section{Summary}
In summary, we introduce the Jones polynomial and the topological twisted super Yang-Mills theory. In fact, 
that the Jones polynomial of a knot, which is the Euler characteristic of Khovanov homology, can be computed 
by counting the solutions of certain gauge theory equations in four-dimension. Witten find that the Hawking-Page 
phase transition in $AdS_{3}$ space is Lee-Yang type. Then we study the phase transition of the topological 
twisted super Yang-Mills field theory. In fact, the partition function is modular form, which is a holomorphic 
function on the upper half plane with finite $N$. But in the large $N$ limit, the function is not analytic. 
Then, by Lee-Yang theory, phase transition can happen.


The $\frac{1}{N}$ expansion of the free energy of Chern-Simons theory~\cite{Neitzke:2004ni} takes the form 
$F=\sum_{g, h} C_{g, h} N^{h} \kappa^{2 g-2+h}=\sum_{g, h} C_{g, h} N^{2-2 g} \lambda^{2 g-2+h}$. However, 
in the large $N$ limit, it is possible to have phase transitions even in finite volume. Further evidence was 
found in that the Wilson lines observable in the topological string theory gives the same knot 
invariants~\cite{Labastida:2000yw}. Then it is natural to study  phase transition in the topological string theory.

\section*{Acknowledgment}
Part of the work was done when Jing Zhou visited the Yau Mathematical Science Center.
Jing Zhou thanks for discussing with Si Li, Hai Lin and Xun Chen. This work is partly supported by the
National Science Foundation of China under Contract Nos. 11775118, 11535005.

\end{document}